**Direct frequency comb saturation spectroscopy with an ultradense tooth spacing of 100 Hz**

D. A. Long,[1,*] A. J. Fleisher,[1] and J. T. Hodges[1]

[1]*Material Measurement Laboratory, National Institute of Standards and Technology,*
*100 Bureau Drive, Gaithersburg, Maryland 20899, USA*

*Corresponding author: David A. Long (david.long@nist.gov; Tel.:(301)-975-3298; Fax: (301)-869-4020).

**Abstract**

Electro-optic frequency combs with tooth spacings as low as 100 Hz were employed to probe electromagnetically induced transparency (EIT) and hyperfine pumping in rubidium and potassium vapor cells. From the potassium EIT transition we were able to determine the ground state hyperfine splitting with a fit uncertainty of 8 Hz. Importantly, because of the mutual coherence between the control and probe beams, which originate from a single laser, features with linewidths several orders-of-magnitude narrower than the laser linewidth could be observed in a multiplexed fashion. This approach removes the need for slow scanning of either a single laser or a traditional mode-locked-laser-based optical frequency comb.

Electro-optic frequency combs allow for an unprecedented degree of control over their parameters, providing for digital control over their span, comb tooth spacing, and central frequency (e.g., see Ref. [1] and the references contained therein). Critically, with the development of waveguide-based electro-optic modulators, the comb tooth spacing is independent of the laser cavity length and can therefore be made essentially arbitrarily narrow. As a result, combs can readily be produced with mode spacings more than six orders of magnitude denser than with a traditional mode-locked-laser-based optical frequency comb. This ultradense mode spacing is ideally suited to the study of narrow sub-Doppler features in atomic and molecular systems.

Here we utilize an electro-optic phase modulator in concert with an arbitrary waveform generator to produce optical frequency combs [2,3] with tooth spacings as small as 100 Hz which are located in the near-infrared. These combs allow for the multiplexed study of saturation spectroscopy whereby the entire spectrum can be acquired in a single interferogram, removing the need for slow spectral interleaving. We have applied this platform to rubidium and potassium transitions in the near-infrared region, probing ultranarrow electromagnetically induced transparency features in buffer gas and paraffin coated cells. This approach enables the interrogation of features that are significantly narrower than 1 kHz and even narrower than the laser linewidth.

The experimental setup was similar to that found in Ref. [3] (see Fig. 1 of that publication for an optical schematic) in which a self-heterodyne configuration was utilized [4,5]. The laser source was a fiber-coupled, external-cavity diode laser (ECDL) which was split into probe and local oscillator legs. We note that two separate ECDLs were used over the course of these measurements, the two lasers were similar with the primary difference being the linewidth, with the first having a reported linewidth of 15 kHz in 10 ms whereas the linewidth for the second was estimated to be 50 kHz to 200 kHz at 5 µs. As will be discussed later, this difference in linewidth did not affect the measurements. The probe path passed through a waveguide-based electro-optic phase modulator (EOM) for comb generation. The EOM was driven by repeated, linear frequency chirps generated by an arbitrary waveform generator (AWG) having a maximum bandwidth of 5 GHz and operating at 12-bit vertical resolution. AWG sampling rates between $1.5 \times 10^9$ samples per second and $1 \times 10^{10}$ samples per second were employed for these measurements. A typical, ultrahigh resolution optical frequency comb (following down conversion into the radiofrequency, RF, domain) can be found in Fig. 1. An acousto-optic frequency shifter was employed to ensure that the positive- and negative-order comb teeth occurred at unique frequencies in the RF. After being launched into free space, the probe beam was expanded to a $1/e^2$ diameter which was varied between 2.2 mm and 5.9 mm over the course of these measurements and made circularly polarized before being sent into an atomic vapor absorption cell.

Two separate absorption cells were used during these measurements. The first was 25 cm long with a diameter of 2.54 cm and contained rubidium vapor at natural isotopic abundance. The walls were coated with paraffin to prevent wall dephasing (see Ref. [6] and the references therein). The second cell was 7.5 cm in length with a diameter of 2.54 cm and contained potassium vapor at natural isotopic abundance as well as 2.67 kPa of argon. Millions of collisions of the absorber with a buffer gas can occur before loss of coherence, thus leading to a dramatic reduction of transit time broadening [7-9]. Stray magnetic fields that affect each cell (measured to be ≤ 86 mG) were reduced by roughly three orders of magnitude by a triply shielded nickel-iron alloy chamber. An insulated box was then placed around this chamber to allow for heating and temperature control. The rubidium cell was utilized at room temperature whereas the potassium cell was heated to 36 °C.

After the probe beam passed through the cell it was returned to linear polarization and was then relaunched into fiber where it was recombined with the local oscillator beam and sent to a high bandwidth detector. For the rubidium measurements a 12 GHz bandwidth detector was used with a noise-equivalent power (NEP) of 24 pW/Hz$^{1/2}$, whereas for the potassium measurements a 1 GHz bandwidth detector was employed with an NEP of 31 pW/Hz$^{1/2}$. The detector signal was amplified and then split into two legs; one of which went to a 4 GHz, 8-bit oscilloscope for data acquisition and the other to a phase-locked servo. The latter signal was fed back to the voltage-controlled oscillator which drove the acousto-optic frequency shifter to allow for long-term coherent averaging by removing phase noise in the laser beam caused by thermal and mechanical fiber fluctuations [10]. The laser wavelength was stabilized with a bandwidth of 100 Hz using a high precision wavelength meter (0.5 MHz optimal resolution). The resulting spectra were then normalized against spectra recorded when the laser was detuned far from the relevant absorption features. These normalized complex-valued transmission spectra ($\tilde{T}$) were then converted to absorbance spectra ($\alpha$) and phase spectra ($\phi$) by the expression $\tilde{T} = \exp\{-(\alpha + i\phi)L/2\}$, where the factor of two accounts for the heterodyne nature of the measurements.

An optical frequency comb spanning 2000 MHz to 3600 MHz with a comb tooth spacing of 200 kHz (i.e., containing 16 000 comb teeth) was used to probe the $D_2$ transition of $^{85}$Rb near 780.2 nm. The unmodulated, carrier frequency of the comb served as the pump (i.e., control) beam to initiate the nonlinear spectroscopy. As can be seen in Fig. 2, the pump can be placed within either the $F''=2$ or $F''=3$ transitions and the corresponding hyperfine levels of either $F''=3$ or $F''=2$ can be probed. This allows for multiplexed observations of hyperfine pumping and electromagnetically induced transparency (EIT). Experimental spectra were simulated using a complex-valued transmission model, allowing for the inclusion of complex-valued etalons [11] as well as velocity selective optical pumping resulting from counterpropagating laser beam reflections [2]. The fit to these spectra yielded measurements of the upper state hyperfine splittings with fit values of 29.372 MHz ($F'$=1 to $F'$=2), 63.29 MHz ($F'$=2 to $F'$=3), and 121.60 MHz ($F'$=3 to $F'$=4) and estimated precisions limited by observed spectral signal-to-noise ratios and by variations in the experimental baseline of 51 kHz ($F'$=1 to $F'$=2), 110 kHz ($F'$=2 to $F'$=3), and 440 kHz ($F'$=3 to $F'$=4).

Because of the digitally controlled nature of these optical frequency combs we can also dramatically reduce the comb tooth spacing required for ultradense comb measurements of the EIT feature (located near ± 3035 MHz in Fig. 2). In Fig. 3 we employed an optical frequency comb containing 5 000 comb teeth spaced by 200 Hz to probe this ultranarrow feature. The shown feature has a full-width at half-maximum of 35 kHz, more than two orders of magnitude narrower than the natural linewidth of 6 MHz [12]. Based upon earlier measurements, the EIT lineshape is expected to have two components, a broader feature whose width is given by transit time broadening and a narrower feature which is limited by either the atomic dark time or the decoherence time due to wall collisions [13]. In Fig. 3, the width of the shown EIT feature is consistent with transit time broadening. Due to the low intensity of the pump beam (0.4 mW/cm$^2$ for the conditions in this figure) the width of the narrow EIT feature is expected to be approximately 200 Hz but have very small contrast [13], thus likely accounting for the lack of observation herein.

To demonstrate our experimental capabilities in the limit where the EIT feature is dramatically narrower than the laser linewidth, we utilized the potassium cell with an argon buffer gas to probe the $^{39}$K $D_1$ transition near 770.1 nm. The optical frequency comb used to interrogate this cell can be seen in Fig. 1, with a comb tooth spacing of only 100 Hz. As can be seen in Fig. 4, this comb can clearly resolve an EIT feature with a width of only 790(20) Hz. Importantly, this width is a factor of 60 to 240 narrower than the estimated laser linewidth. This sub-laser-linewidth resolution is enabled by the mutual coherence between the pump and probe beams, thus allowing for a significant reduction of the apparent frequency jitter. Further, the use of the second ECDL with the far narrower linewidth led to similar results, indicating that laser linewidth is not a limiting factor in the self-heterodyne frequency resolution.

We have shown that a self-heterodyne configuration of electro-optic frequency combs enables multiplexed direct frequency comb spectroscopy with unprecedented resolution. Unlike measurement techniques that employ traditional mode-locked-laser-based optical frequency combs, no slow interleaving of combs with relatively coarse tooth spacing is required to achieve high resolution. The present approach exploits the high mutual coherence of the pump (control) and probe fields to enable the efficient preparation and highly resolved observation of dressed states. These attributes make the present method attractive for high-precision frequency metrology in atomic and molecular systems.

**Acknowledgements**

We would like to acknowledge David F. Plusquellic (Physical Measurement Laboratory, National Institute of Standards and Technology) for helpful discussions and for providing software to interface with the arbitrary waveform generator.

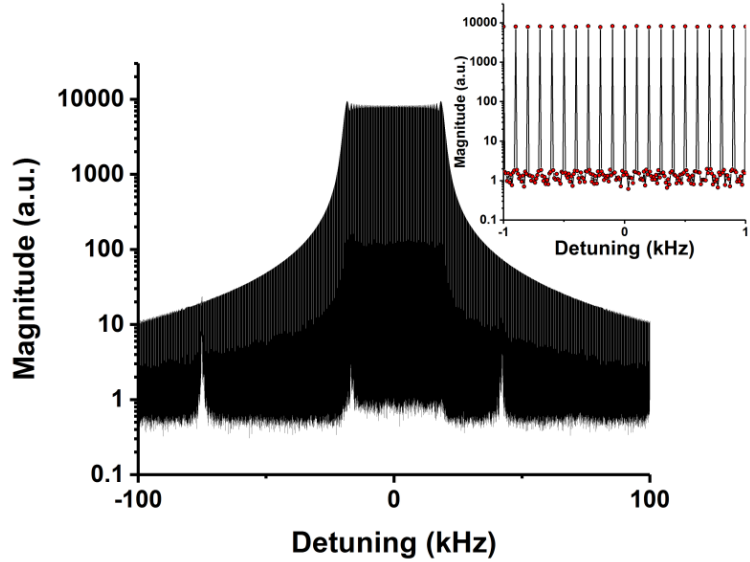

FIG. 1. Typical ultrahigh resolution optical frequency comb containing 400 comb teeth spaced at 100 Hz. The shown self-heterodyne spectrum is the average of the ten off-resonance frequency domain traces shown in Figure 4. Each component trace was the magnitude of the Fourier transform of one thousand coherently averaged interferograms which contained $50\times10^6$ samples recorded at $500\times10^6$ samples per second. The inset shows a zoomed-in-view of the optical frequency comb. Note that the comb teeth are resolution bandwidth limited. The relative standard deviation of magnitudes of the twenty-one comb teeth shown in the inset is only 1.7%.

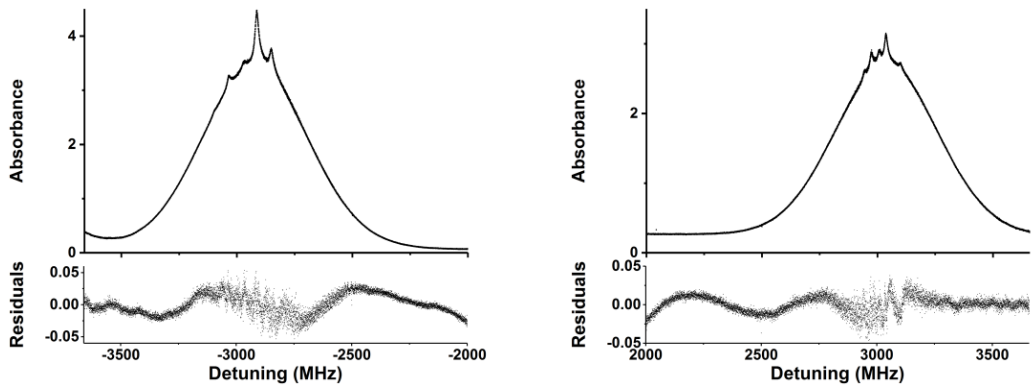

FIG. 2. Probe comb spectra and fit residuals of the $^{85}$Rb $D_2$ $F''=3$ (left panel) and $F''=2$ (right panel) transitions recorded in a room temperature rubidium vapor cell containing a paraffin wall coating. The pump (i.e., carrier tone) is located at zero detuning, which saturates the $F''=2$ (left panel) and $F''=3$ (right panel) transitions. A series of hyperfine pumping transitions can be observed which are separated by the upper state hyperfine splittings. These spectra were recorded with a 200 kHz-spaced comb. For each spectrum, five thousand interferograms were coherently averaged in the time domain with each containing $4\times10^7$ samples recorded at $2\times10^{10}$ samples per second.

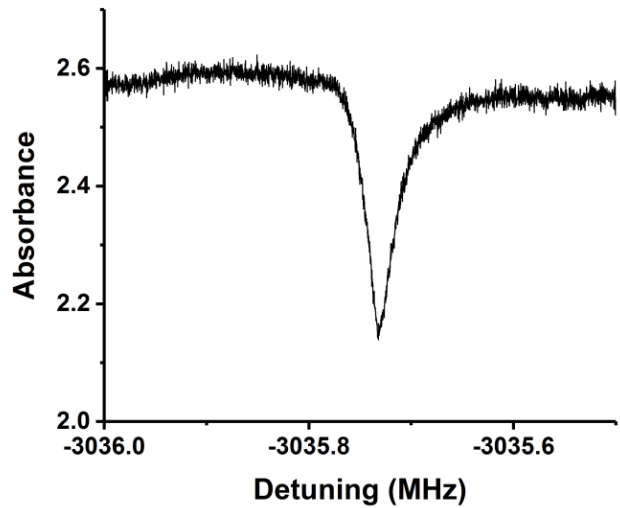

FIG. 3. Electromagnetically induced transparency spectrum for the $^{85}$Rb $D_2$ transition recorded in a room temperature rubidium vapor cell containing a paraffin wall coating. A comb tooth spacing of 200 Hz was employed. Two hundred interferograms (each containing $1\times10^8$ samples recorded at $1\times10^{10}$ samples per second) were coherently averaged in the time domain.

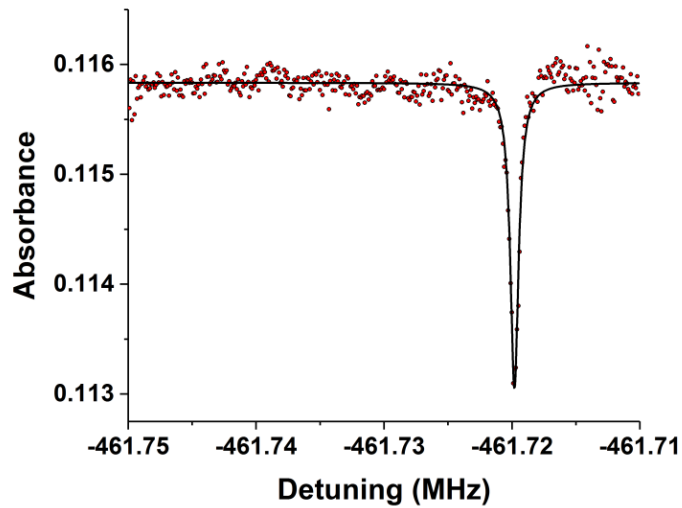

FIG. 4. Electromagnetically induced transparency spectrum for the $^{39}$K $D_1$ transition and corresponding Lorentzian fit in a potassium vapor cell. The cell contained 2.67 kPa of argon buffer gas and was held at 36 °C. The shown spectrum was recorded with a comb tooth spacing of 100 Hz. One thousand interferograms were coherently averaged in the time domain before being Fourier transformed and normalized to produce a spectrum. Each interferogram contained $5\times10^7$ samples recorded at $5\times10^8$ samples per second. Ten of these spectra were then averaged to produce the shown image. The shown Lorentzian fit yielded a full-width half-maximum of 790(20) Hz with a fit uncertainty on the line center of only 8 Hz.

## References

[1]  V. Torres-Company and A. M. Weiner, Laser Photon. Rev. **8**, 368 (2014).

[2]     D. A. Long, A. J. Fleisher, D. F. Plusquellic, and J. T. Hodges, Phys. Rev. A **94**, 061801 (2016).
[3]     D. A. Long, A. J. Fleisher, D. F. Plusquellic, and J. T. Hodges, Opt. Lett. **42**, 4430 (2017).
[4]     N. B. Hébert, V. Michaud-Belleau, J. D. Anstie, J. D. Deschênes, A. N. Luiten, and J. Genest, Opt. Express **23**, 27806 (2015).
[5]     Y. Bao, X. Yi, Z. Li, Q. Chen, J. Li, X. Fan, and X. Zhang, Light Sci. Appl. **4**, e300 (2015).
[6]     K. Nasyrov, S. Gozzini, A. Lucchesini, C. Marinelli, S. Gateva, S. Cartaleva, and L. Marmugi, Phys. Rev. A **92**, 10, 043803 (2015).
[7]     S. Brandt, A. Nagel, R. Wynands, and D. Meschede, Phys. Rev. A **56**, R1063 (1997).
[8]     S. Ezekiel, S. P. Smith, M. S. Shahriar, and P. R. Hemmer, J. Lightwave Technol. **13**, 1189 (1995).
[9]     D. E. Nikonov, U. W. Rathe, M. O. Scully, S. Y. Zhu, E. S. Fry, X. F. Li, G. G. Padmabandu, and M. Fleischhauer, Quantum Opt. **6**, 245 (1994).
[10]    A. J. Fleisher, D. A. Long, Z. D. Reed, J. T. Hodges, and D. F. Plusquellic, Opt. Express **24**, 10424 (2016).
[11]    A. J. Fleisher, D. A. Long, and J. T. Hodges, J. Mol. Spectrosc. **352**, 26 (2018).
[12]    U. Volz and H. Schmoranzer, Phys. Scr. **T65**, 48 (1996).
[13]    M. Klein, M. Hohensee, D. F. Phillips, and R. L. Walsworth, Phys. Rev. A **83**, 013826 (2011).